\documentclass[prl,nofootinbib,superscriptaddress,twocolumn]{revtex4}
\usepackage[a4paper,left=1.5cm,right=1.5cm,top=3cm,bottom=3cm]{geometry}

\usepackage{amssymb,amsmath,amsfonts}
\usepackage[dvipsnames]{xcolor}
\usepackage{graphicx}
\usepackage{longtable}
\usepackage{verbatim}
\usepackage{color}
\usepackage{mdframed}
\usepackage{soul}
\usepackage{amsfonts,amssymb,mathrsfs,amsmath,esint}
\usepackage{slashed, cancel}
\usepackage{framed}
\usepackage{mdframed}
\usepackage{simplewick} 
\allowdisplaybreaks

\usepackage{latexsym}
\usepackage{graphicx}
\graphicspath{{./Figures/}}
\usepackage[dvipsnames]{xcolor}
\usepackage{booktabs}
\usepackage{datetime}
\newdateformat{mydate}{\THEDAY{ }\monthname[\THEMONTH]{ }\THEYEAR}

\usepackage{tikz}
\usepackage{color}
\usepackage{framed}
\usepackage{hyperref}
\hypersetup{colorlinks, citecolor=bluscuro, linkcolor=black, urlcolor=bluscuro}
\definecolor{rossos}{cmyk}{0,1,1,0.55}
\definecolor{bluscuro}{rgb}{0.15, 0.2, .85}
\definecolor{bluchiaro}{cmyk}{1,.3,0.,0.1}

\graphicspath{{./Figures/}}

\newcommand{\be}{\begin{equation}}
\newcommand{\ee}{\end{equation}}
\newcommand{\bea}{\begin{eqnarray}}
\newcommand{\eea}{\end{eqnarray}}
\newcommand{\beq}{\begin{equation}}
\newcommand{\eeq}{\end{equation}}
\def\beqa{\begin{eqnarray}}

\def\CMB{{\text{\tiny CMB}}}

\def\GW{{\text{\tiny GW}}}

\def\eeqa{\end{eqnarray}}

\def\lsim{\mathrel{\rlap{\lower4pt\hbox{\hskip0.5pt$\sim$}}
    \raise1pt\hbox{$<$}}}         
\def\gsim{\mathrel{\rlap{\lower4pt\hbox{\hskip0.5pt$\sim$}}
    \raise1pt\hbox{$>$}}}         

\newcommand{\eff}{\rm eff}

\usepackage[normalem]{ulem}
\usepackage{soul}

\begin{document}

\title{The Imprint of Relativistic Particles on the Anisotropies of the \\Stochastic Gravitational-Wave Background}

\author{L. Valbusa Dall'Armi}
\address{Dipartimento di Fisica e Astronomia ``G. Galilei",
Universit\`a degli Studi di Padova, via Marzolo 8, I-35131 Padova, Italy}

\author{A. Ricciardone}

\address{INFN, Sezione di Padova,
via Marzolo 8, I-35131 Padova, Italy}

\author{N. Bartolo}
\address{Dipartimento di Fisica e Astronomia ``G. Galilei",
Universit\`a degli Studi di Padova, via Marzolo 8, I-35131 Padova, Italy}

\address{INFN, Sezione di Padova,
via Marzolo 8, I-35131 Padova, Italy}

\address{INAF - Osservatorio Astronomico di Padova, Vicolo dell'Osservatorio 5, I-35122 Padova, Italy}

\author{D. Bertacca}
\address{Dipartimento di Fisica e Astronomia ``G. Galilei",
Universit\`a degli Studi di Padova, via Marzolo 8, I-35131 Padova, Italy}

\address{INFN, Sezione di Padova,
via Marzolo 8, I-35131 Padova, Italy}

\author{S. Matarrese}
\address{Dipartimento di Fisica e Astronomia ``G. Galilei",
Universit\`a degli Studi di Padova, via Marzolo 8, I-35131 Padova, Italy}

\address{INFN, Sezione di Padova,
via Marzolo 8, I-35131 Padova, Italy}

\address{INAF - Osservatorio Astronomico di Padova, Vicolo dell'Osservatorio 5, I-35122 Padova, Italy}

\address{Gran Sasso Science Institute, Viale F. Crispi 7, I-67100 L'Aquila, Italy}

\date{\today}

\begin{abstract}
\noindent
The Stochastic Gravitational-Wave Background (SGWB) is expected to be a key observable for Gravitational-Wave (GW) interferometry. Its detection will open a new window on early Universe cosmology, on the astrophysics of compact objects and, as shown in this Letter, on the particle physics content of the Universe. In this Letter we show that, besides their effects on the Cosmic Microwave Background (CMB) and on Large Scale Structure (LSS), relativistic particles in the early Universe leave a clear imprint on the anisotropies of the SGWB. In particular we show that a change in the number of decoupled relativistic particles shifts the angular power spectrum of the SGWB, as both the Sachs-Wolfe (SW) and the Integrated Sachs-Wolfe (ISW) terms are affected. Being very large-angle effects, these lead to new testable predictions for future GW interferometers. 

\end{abstract}

\maketitle

\paragraph{Introduction.}

Future GW interferometers \cite{Audley:2017drz, Kawamura:2006up, Evans:2016mbw, Sathyaprakash:2011bh, Maggiore:2019uih} will probe the SGWB from late-time unresolved astrophysical sources and early Universe cosmological sources (see e.g. \cite{Regimbau:2011rp, Caprini:2018mtu} for reviews). Besides the important information about astrophysics and cosmology~\cite{Maggiore:1999vm, Guzzetti:2016mkm,  Bartolo:2016ami}, such detections will allow to extract important information about particle physics within or beyond the Standard Model (SM) \cite{Barausse:2020rsu}. For many years the interplay between cosmology and particle physics has been pursued vigorously; one  example being the constraints on the effective number of neutrinos $N_\nu$ and on neutrino masses, which have been largely investigated using CMB and LSS data~\cite{Crotty:2003th, Hannestad:2005jj, Cuoco:2003cu, Lesgourgues:2006nd, Aghanim:2018eyx}. In this Letter we will show that the cosmological SGWB will offer a new powerful tool to constrain the abundance of relativistic species in the the early Universe.

As recently shown~\cite{Contaldi:2016koz, Bartolo:2019oiq, Bartolo:2019yeu}, using a  Boltzmann equation approach, it is possible to characterize angular anisotropies of the GW energy density, thus providing an important tool to disentangle the different cosmological and astrophysical contributions to the SGWB \cite{Bertacca:2019fnt, Caprini:2019pxz,Pieroni:2020rob}. Anisotropies in the cosmological background are imprinted both at its production, and by GW  propagation through the large-scale scalar~\cite{Alba:2015cms,Contaldi:2016koz, Bartolo:2019oiq, Bartolo:2019yeu} and tensor perturbations of the Universe~\cite{Contaldi:2016koz, Bartolo:2019oiq, Bartolo:2019yeu}. In considering the SGWB there is a crucial difference w.r.t the CMB: while CMB temperature anisotropies are generated at the last scattering surface~\cite{Peebles:1970ag, Bond:1983hb}, the Universe is transparent to GWs at all energies below the Planck scale~\cite{Misner:1974qy, Garoffolo:2019mna}. Therefore, the SGWB provides a snapshot of the Universe  at the epoch of its production, and its anisotropies retain precious information about the primordial Universe, the mechanisms for GW formation and the presence of extra particle species in such an era.

In this Letter, we will focus on the  impact of the effective number of relativistic degrees of freedom $N_{\rm eff}$ on the anisotropies of the stochastic background of cosmological origin generated from the propagation of GWs in the perturbed Universe. Although this effect is present also for different cosmological sources of GWs (e.g. phase transition, cosmic strings, preheating, etc), we will consider the SGWB generated during inflation in the early Universe.  The number of relativistic degrees of freedom will have a direct impact on the angular power-spectrum of the SGWB, on scales accessible to GW interferometers, mainly through the Sachs-Wolfe (SW) and Integrated Sachs-Wolfe (ISW) effects.   
The SW effect is due to the gravitational redshift that affects particles when they started their free streaming across the Universe. Due to cosmological inhomogeneities, gravitons which come from different directions started their free streaming under the action of different gravitational potentials, thus inhomogeneities in the metric generate anisotropies in the angular power-spectrum. The ISW effect is also caused by gravitational redshift, but integrated in time: when a particle is crossing a perturbation, if the perturbation changes in time the particle gains or looses energy. 

The overall effect of increasing the number of relativistic species on the CMB angular power-spectrum is a horizontal and a vertical shift of the peak positions, respectively: increasing the value of $N_{\eff}$ the height of the first peak is enhanced and the positions of the acoustic peaks are shifted to higher multipoles~\cite{Hu:1995em,  Calabrese:2011hg}. The first effect derives from the change of the epoch of matter-radiation equality. In fact, by increasing (decreasing) the value of $N_{\eff}$, the matter-radiation equality occurs later (earlier). This brings to an enhancement of the first peak, due to the early ISW effect (and the radiation driving effect), 
in which fluctuations of the corresponding scale, having crossed the sound horizon in the radiation-dominated epoch, are boosted by the decay of the gravitational potential. A similar effect comes from a decrease (increase) of the matter energy density \cite{Hu:1995em}. The second effect, i.e. shift of the positions of the peaks, is again related to the change of the matter-radiation equality, through the change of $N_{\eff}$, and in particular on the effect on the sound horizon (at the recombination epoch) which becomes smaller when $N_{\eff}$ increases \cite{Kolb:1990vq,Dodelson:2003ft}.\\
Another well-known effect given by decoupled relativistic species, in particular neutrinos, on the CMB  angular power-spectrum, is the damping due to their anisotropic stress of the amplitude of the GW spectrum by~$35\%$~\cite{Weinberg:2003ur,  Dicus:2005rh, Stefanek:2012hj}. Such an effect is quite enhanced in the frequency region between $10^{-16}$ Hz and   $10^{-10}$ Hz, while is less significant below $10^{-16}$ Hz, since this frequency region probes the Universe when it was matter-dominated~\cite{Weinberg:2003ur, Watanabe:2006qe, Mangilli:2008bw}.

For the SGWB, we know that gravitons decoupled long before the end of inflation, so the time at which the SW effect is generated is much earlier and the integration time for the ISW effect is much longer. In this Letter we compute the imprints of relativistic particles on the SGWB angular power-spectrum: in particular, as far as the SW term is concerned, the effect is mainly due to the fact that gravitons decoupled at much larger energy scales (w.r.t. CMB photons) and the consequence of adding new species is to suppress its contribution.
For the ISW effect we know that it is determined by the variation of the gravitational potentials $\Phi$ and $\Psi$ from graviton decoupling until the present epoch. As we know from the CMB \cite{Dodelson:2003ft}, there are two different ISW contributions: the Early ISW, generated by the time-variation of the modes when they cross the equality epoch, and the Late ISW, due to the drastic changes of $\Phi$ and $\Psi$ in the dark-energy-dominated era. The first term is responsible for the enhancement of the first acoustic peak (together with the radiation driving effect), while the latter increases by increasing the angular scale, i.e. by reducing $\ell$. There is a key difference between the CMB and the SGWB: photons decoupled at last scattering, during the matter-dominated era, while gravitons decoupled long before (as detailed below), therefore the integration for computing the ISW effect for the gravitons started a long earlier. This essentially does not modify the $\ell$-ISW, while it changes considerably the  early ISW. 


Following~\cite{Dodelson:2003ft,Hu:1995em,Contaldi:2016koz, Bartolo:2019oiq, Bartolo:2019yeu} we start by defining the distribution function $f$ for the gravitons and  we write down the Boltzmann equation for $f$ in a perturbed spatially flat Friedmann-Lema\^itre-Robertson-Walker (FLRW) metric, accounting that graviton move along null geodesics defined by the background metric (which should be understood including large scale perturbations); we consider gravitons as collisionless particles, under the assumption that they decouple at early times. We include the effect of extra relativistic degrees of freedom on the scalar gravitational potentials through their contribution to the anisotropic stress, and then we quantify the impact on the angular SGWB power spectrum. We will briefly comment also on the effect on the tensor contribution to the SGWB spectrum, even if, as we will see, such effects will be important at much smaller scales. 

\noindent
\vskip 0.3cm
\noindent

\paragraph{Boltzmann equation for GWs.}
\noindent
To study the effect of new particle species on SGWB anisotropies we start from the Boltzmann equation for the gravitons $\mathcal{L}[f] = \mathcal{C}[f(\lambda)] + \mathcal{I}[f(\lambda)]$,
where $\mathcal{L}\equiv d/ d\lambda$ is the Liouville term, $\mathcal{C}$ and $ \mathcal{I}$ account respectively for the collisions of GWs along their path, and for their emissivity from cosmological and astrophysical sources \cite{Contaldi:2016koz}. The collisions among  GWs can be disregarded since they affect the distribution at higher orders 
(in a series expansion in the gravitational strength 
$1/M_{\rm P}$, see \cite{Bartolo:2018igk}). The emissivity can be related to some astrophysical processes (such as merging of compact objects) in the late Universe, as well as cosmological processes, so we treat the emissivity term as an initial condition on the GW distribution.  The background metric on which our gravitons propagate is defined by 
\begin{equation}
ds^2=a^2(\eta)\left[
-e^{2\Phi} d\eta^2+(e^{-2\Psi}\delta_{ij}+ h_{ij}) 
dx^i dx^j\right]\, ,
\label{metric}
\end{equation}
where $a ( \eta )$ is the scale-factor,  $\eta$ is conformal time and we consider only scalar ($\Phi$ and $\Psi$) and tensor ($h_{ij}$, taken to be transverse and traceless) perturbations in the so-called Poisson gauge. As we will show in the following, the effect of new particle species on the SGWB will be transferred via their effect on the scalar potentials. 

The Boltzmann equation for the SGWB of cosmological origin, can be computed in a similar way to what is done for the CMB \cite{Dodelson:2003ft, Bartolo:2006fj, Bartolo:2006cu}. Keeping only the terms up to first order in the perturbations, it reads~\cite{Bartolo:2019oiq, Bartolo:2019yeu}
\begin{equation}
\frac{\partial f}{\partial \eta} + n^i \, \frac{\partial f}{\partial x^i} + \left[ \frac{\partial \Psi}{\partial \eta} - n^i \, \frac{\partial \Phi}{\partial x^i} + \frac{1}{2} \, n^i \, n^j \, \frac{\partial h_{ij}}{\partial \eta} \right] q \, \frac{\partial f}{\partial q} = 0 \, ,
\label{Boltzmann-1st}
\end{equation}
where ${\hat n} \equiv {\hat p}$ is the direction of motion of the GWs, while $q \equiv \vert \vec{p} \vert a$ is the comoving momentum.
The first two terms encode free streaming, that is the propagation of perturbations on all scales. The third term causes the red-shifting of gravitons, including SW, ISW and Rees-Sciama (RS) effects~\cite{Hu:1995em}. As we will see in the next section, through the scalar potentials, both the SW and the ISW, are affected by the presence of extra-particle species after inflation, and so is the SGWB. It is convenient to rescale the perturbed part of the distribution function using the following redefinition
$\delta f \equiv  - q \, (\partial {\bar f}/\partial q) \, \Gamma \left( \eta ,\, \vec{x} ,\, q ,\, {\hat n} \right), $
%
$\bar{f}$ being the homogeneous and isotropic contribution, and the Boltzmann equation in Fourier space, eq. (\ref{Boltzmann-1st}), becomes 
\begin{equation}
\Gamma'+ i \, k \, \mu\, \Gamma = S (\eta, \vec{k}, {\hat n})  \, ,
\label{Boltfirstgamma1}
\end{equation}
where the source function is $ S  = \Psi' - i k \, \mu \, \Phi -  n^i n^j  \, h_{ij}' /2$ (primes denoting differentiation w.r.t. conformal time and $\mu\equiv \hat{k}\cdot {\hat n}$).
The quantity $\Gamma$ can be immediately related to the perturbation of the GW energy density, $\rho_\GW \equiv \int d^3 p \, p \, f $. It is customary to parameterize the GW energy density measured at time $\eta$ and location $\vec{x}$ in terms of its fractional contribution $\Omega_\GW$, through 
$\rho_\GW \left( \eta ,\, \vec{x} \right) \equiv \rho_{\rm crit} \int d  \ln q \, \Omega_\GW \left( \eta ,\, \vec{x} ,\, q \right) $\;, 
where $\rho_{\rm crit} = 3 H^2 M_{\rm P}^2$ is the Universe critical energy density, and $H$ the Hubble rate. Since we are interested in its inhomogeneous and anisotropic component, we 
allowed $\Omega_\GW$ to depend on space. We account for the anisotropic dependence by defining $\omega_\GW$ through $\Omega_\GW = \int d^2 {\hat n} \, \omega_\GW ( \eta ,\, \vec{x} ,\, q ,\, {\hat n} )/4\pi$, and by introducing the density contrast $\delta_\GW \equiv  \delta \omega_\GW ( \eta ,\, \vec{x} ,\, q ,\, {\hat n} )/\bar \omega_\GW ( \eta ,\, q ) $. Using the $\Gamma$ definition, introduced in \cite{Bartolo:2019oiq, Bartolo:2019yeu}, one then finds 
\begin{equation}
\delta_\GW  = \left[ 4 -  \frac{\partial \ln \, {\bar \Omega}_\GW \left( \eta ,\, q \right)}{\partial \ln \, q}  \right] \, \Gamma \left( \eta ,\, \vec{x} ,\, q ,\, {\hat n} \right) \,, 
\label{delta-Gamma}
\end{equation} 
with ${\bar \Omega}_\GW$ the homogeneous, isotropic component of $\Omega_\GW$. 

\vskip 0.1cm
As shown in~\cite{Watanabe:2006qe}, $\bar{\Omega}_\GW$ is sensitive to the evolution of the relativistic degrees of freedom $g_*$ before matter-radiation equality. The relativistic degrees of freedom can be expressed in terms of the photon temperature $T$, the intrinsic degrees of freedom and temperature of the various particle species $g_\alpha$ and $T_\alpha$ as
\begin{equation}
g_*(T)=\sum_{\alpha,BE}g_\alpha \Bigl(\frac{T_\alpha}{T}\Bigl)^4+\frac{7}{8}\sum_{\alpha,FD}g_\alpha \Bigl(\frac{T_\alpha}{T}\Bigl)^4\,.
\end{equation} 
With $BE$ we mean integer spin particles which follow Bose-Einstein statistics, while with $FD$ we identify semi-integer spin particles which follow Fermi-Dirac statistics. The energy density of relativistic particles can be written then in terms of $g_*$ as
$\rho(T)=\pi^2/30 g_*(T)T^4.$
From the end of inflation until the present epoch, the temperature of the different particle species decreases, and many of them become non-relativistic, $T_\alpha\lesssim m_\alpha$, giving no more contribution to $g_*$, which changes from $g_*(T\gtrsim 10^4\, {\rm MeV})\simeq 106$, when all the SM particles contribute, to $g_*(T\lesssim 0.1 \, {\rm MeV}) = 3.36$, when only photons and relativistic neutrinos contribute~\cite{Kolb:1990vq, Lesgourges}. 

\vskip 0.3cm
\paragraph{ Effects on the scalar perturbations}
\noindent
The main role of relativistic particles is played on the ``scalar'' part of the anisotropic stress
\begin{equation}
\label{eq: nuscalareq}
k^2(\Phi-\Psi)=-32\pi G a^2 \rho_r \mathcal{N}_{2},
\end{equation}
where $\mathcal{N}_{2}$ is the quadrupole moment generated by the relativistic particles.
The fractional energy density of  decoupled relativistic particles can be described in terms of degrees of freedom as 
\begin{equation}
f_{\rm dec}(\eta_i)\equiv g_*^{\rm dec}(T_i)/g_*(T_i)\, ,
\end{equation}
 where $g_*^{\rm dec}(T_i)$ are the relativistic degrees of freedom of decoupled particles evaluated at temperature $T_i$ at the end of inflation, corresponding to conformal time $\eta_i$. This influences the initial conditions for the scalar metric perturbations at the end of inflation $\eta_i$~\cite{Bashinsky:2003tk,{Ma:1995ey}}: 
 \begin{equation}
 \label{psi_initial}
 \Psi(\eta_i,k)=\Bigl(1+\frac{2}{5}f_{\rm dec}(\eta_i)\Bigl)\Phi(\eta_i,k),
 \end{equation}
 where the initial value of $\Phi$ is related to the value of the gauge-invariant curvature perturbation $\zeta$ of comoving spatial hyper-surfaces at the end of inflation, 
 $\zeta(\eta_{i},k)=\zeta_I(k)$, 
 \begin{equation}
 \label{phi_initial}
 \Phi(\eta_i,k)=-\frac{2}{3}\Bigl(1+\frac{4}{15}f_{\rm dec}(\eta_i)\Bigl)^{-1}\zeta_I(k).
 \end{equation}
 The fractional energy density of  decoupled relativistic  particles varies since $\eta_i$ down to temperatures around $0.1 \,$MeV, when it reaches a constant value which depends on the chosen $N_{\eff}$; for instance for 3 light neutrino species it corresponds to  $f_{\rm dec}(\eta_{T<0.1\,{\rm MeV}})=0.4$ (different evolutions of $f_{\rm dec}(\eta)$ for different particle candidates are shown for instance in~\cite{DePorzio:2020wcz}). In this interval $\Phi$ and $\Psi$ evolve following eqs. \eqref{psi_initial} and \eqref{phi_initial} for different $f_{\rm dec}(\eta)$ values. At lower temperatures it is well known that small-scale modes start decaying and oscillating once they cross the Hubble horizon before matter-radiation equality~\cite{Dodelson:2003ft}. On the other hand, large-scale modes change a little bit around the time of matter-radiation equality, they remain constant during the matter-dominated epoch and then start decaying during the recent dark energy-dominated era~\cite{Hu:1996vq}. Modes with $k\approx k_{\rm eq}$ present an intermediate behaviour between large and small scales: they have rather large variations both around the matter-radiation equality and during the dark-energy-dominated era~\cite{Hu:1994uz}. Until $\eta \gtrsim \eta_{\rm eq}$,  decoupled relativistic  particles make a substantial contribution to the total energy density and eq. \eqref{eq: nuscalareq} shows that $\Phi$ and $\Psi$ evolve differently. For $\eta \gg \eta_{\rm eq}$ no more species contribute considerably to the anisotropic stress and $\Phi$ and $\Psi$ become approximately equal because the Universe is matter dominated. 

\vskip 0.3cm
\paragraph{Correlators of GW anisotropies and extra species contribution.}
\noindent
Following the treatment adopted for CMB anisotropies,  we expand the solution in spherical harmonics, $\Gamma ( {\hat n} ) = \sum_\ell \sum_{m=-\ell}^\ell \Gamma_{\ell m} \, Y_{\ell m} ( {\hat n} )$, where $Y_{\ell m}(\hat{n})$ are spherical harmonics and ${\hat n}$ is the direction of the GW trajectory to the detector. We focus on two contributions, even though, as shown in \cite{Bartolo:2019oiq,Bartolo:2019yeu} there are three contributions to the anisotropies (the third contribution being an intrinsic initial perturbation of the distribution function that is not relevant here for our purposes). There is a first contribution due to the scalar sources in eq. (\ref{Boltfirstgamma1}) 
\begin{eqnarray} 
\frac{\Gamma_{\ell m,S}}{4\pi  \left( - i \right)^{\ell}} &=&  \,  
\int \frac{d^3 k}{\left( 2 \pi \right)^3} \, 
\zeta_I ( k) 
Y_{\ell m}^*( {\hat k}) \, {\cal T}_\ell^{(0)} (  k ,\, \eta_0 ,\, \eta_{i} ),\label{Glm,S}
\end{eqnarray} 
where the scalar transfer function ${\cal T}_\ell^{(0)}$ is the sum of a SW term, similar to CMB photons,  $T_\Phi ( \eta_{i} , k ) \, j_\ell [ {k ( \eta_0 - \eta_{i} )} ]$, plus an ISW term, $ \int_{\eta_{i}}^{\eta_0} d \eta' \, [  {T_\Psi'  ( \eta , k ) +  T_\Phi'  ( \eta , k ) }] \, j_\ell  [ {k ( \eta - \eta_{i} )}]$ (where $j_{\ell}$ are the spherical Bessel functions of order $\ell$ and $\eta_0$ is the conformal time at the present epoch). It is important to notice that in this case $\eta_i$ corresponds to the time at which gravitons decoupled (the end of inflation). In fact, even if gravitational interactions decoupled around the Planck energy scale, the SGWB has been produced after the Planck epoch, i.e. during inflation, thus we can state that the cosmological GWs decoupled at the end of inflation, because at that time they started their free streaming. For the CMB the situation is different: the initial integration time corresponds to recombination, $T_{\rm rec}\simeq 0.3\, {\rm eV}$.\\
Then there is a second contribution $\Gamma_{\ell m,T}$ due to the tensor modes in eq. (\ref{Boltfirstgamma1}), and it is formally analog to eq. (\ref{Glm,S}), with the product $\zeta_I(k) Y_{\ell m}^*(\hat{k})$ replaced by the combination $\sum_{\lambda= \pm2}  {\hat \xi}_\lambda ( \vec{k} ) \, _{- \lambda}Y_{\ell m}^* ( \hat{k} ) $, involving spin-2 spherical harmonics, and with the scalar transfer function replaced by the tensor one ${\cal T}_\ell^{(\pm 2)} ( k ,\, \eta_0 ,\, \eta_{\rm i} ) $, given by 
\be
{\cal T}_\ell^{(\pm 2)}  =\frac{1}{4}  \sqrt{\frac{\left(\ell +2 \right)!}{\left( \ell - 2 \right)!}}  \int_{\eta_{\rm i}}^{\eta_0} \!\!\!\! d \eta \,  h' \left( \eta ,\, k \right)   \frac{j_\ell \left[ k \left( \eta_0 - \eta \right) \right]}{ k^2 \left( \eta_0 - \eta \right)^2},
\ee
where we have introduced a parametrization for the tensor perturbations in terms of their primordial amplitudes $\hat{\xi}_\lambda$, their polarization tensors $e_{ij,\lambda}$ and their transfer function $h$, $h_{ij}(\eta,k)=\sum_{\lambda=\pm2}h(\eta,k)e_{ij,\lambda}(\hat{k})\hat{\xi}_\lambda(\vec{k})$.
 
Therefore the SGWB angular power spectrum reads $\langle \Gamma_{\ell m}   \Gamma_{\ell' m'}^*  \rangle \equiv \delta_{\ell \ell'} \, \delta_{mm'} \, {\widetilde C}_\ell =  \delta_{\ell \ell'} \, \delta_{mm'} [  {\widetilde C}_{\ell,S} + {\widetilde C}_{\ell,T} ] $, where we denote the correlators with a tilde to distinguish them from the CMB case. Focusing only on the scalar contribution to the angular 
power-spectrum we have 
\begin{equation}
\label{coefficienti}
\begin{split}
\frac{\tilde{C}_{\ell,S}(\eta_0)}{4\pi}=&\int \frac{dk}{k}P^{(0)}(k)\biggl\{T_\Phi(\eta_i,k)j_\ell[k(\eta_0-\eta_i)]+\\&\hspace{.01em}+\int_{\eta_i}^{\eta_0} d\eta \Bigl[T_\Phi^\prime(\eta,k)+T_\Psi^\prime(\eta,k)\Bigl] j_\ell[k(\eta_0-\eta)]\biggl\}^2\, ,
\end{split}
\end{equation}
where $P^{(0)}(k)$ is the primordial scalar power-spectrum. 
In the following, we are going to quantify the effect of the extra relativistic species on such terms, which dominate on large scales, and as such they are the ones which can be probed by GW interferometers due to their limited angular resolution \cite{Taruya:2005yf, Taruya:2006kqa,Baker:2019ync,Contaldi:2020rht}. 
\vskip 0.1cm

\paragraph{Sachs-Wolfe effect and relativistic particle species. }

\begin{figure}[t!]
\centering 
\includegraphics[width=0.5\textwidth]{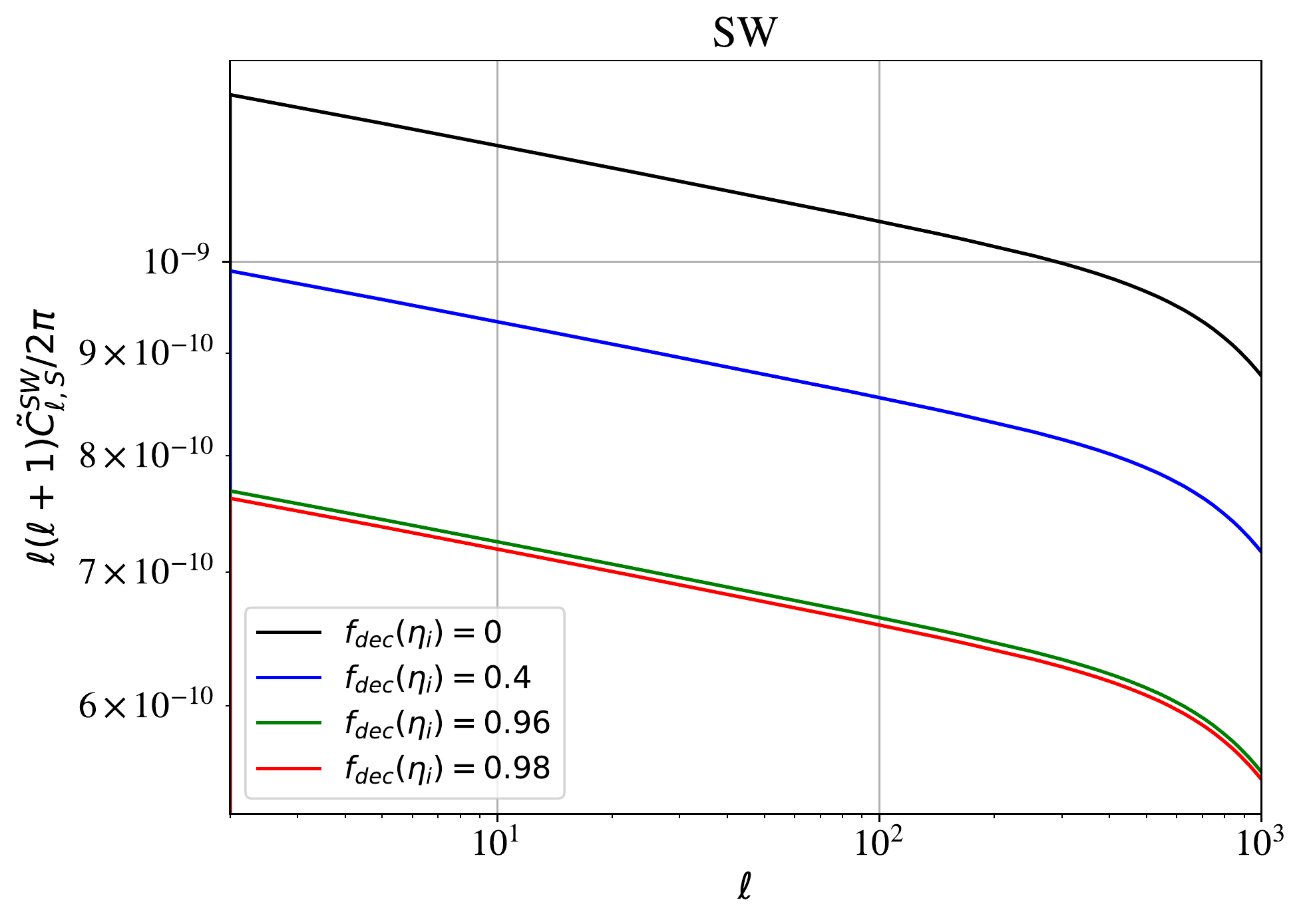}
\vskip -0.4cm
\caption{\it SW contribution to the angular power-spectrum of the SGWB. We can see that by increasing $f_{\rm dec}(\eta_i)$ we are decreasing more and more the amplitude of the angular power-spectrum. For values of $f_{\rm dec}(\eta_i)$ close to 1 we observe a saturation.}
\label{SW_figure}
\end{figure}

Similarly to the CMB case, at large angular scales the dominant term of the scalar contribution to the angular power-spectrum is the SW one. Taking into account the initial time $\eta_i$ for the SGWB case, we modified the public code CLASS for the computation of CMB anisotropies \cite{Lesgourgues:2011re} adapting it to the SGWB.  In Figure~\ref{SW_figure} we have plotted $\tilde{C}_{\ell,S}$, showing how different values of $f_{\rm dec}(\eta_i)$ (and thus different choices for the end of inflation energy scale and implicitly for the number of relativistic particles present at that time) affect differently the spectra. In the absence of a specific particle physics model for describing the decoupled relativistic species at $\eta_i$, we have varied $f_{\rm dec}(\eta_i)$ over all its domain, between 0 and 1, in this way we have determined the maximum and the minimum SW effect (values close to 1 are only considered for illustrative purposes; as such a large fraction is not physically achievable). We can also give a simple analytic estimate of the SW contribution starting from eq. \eqref{phi_initial}. Considering that the effect is generated by particles which are relativistic at their decoupling ($T_{dec} > m$), 
a simple estimate of the damping at low $\ell$ is given by
\begin{equation}
\begin{split}
\frac{\tilde{C}_{\ell,S}^{SW}(\eta_0)}{4\pi}=&\frac{4}{9}\Bigl[1+\frac{4}{15}f_{\rm dec}(\eta_i)\Bigl]^{-2}\times\\
&\times\int \frac{dk}{k}P^{(0)}(k)j^2_\ell[k(\eta_0-\eta_i)]. 
\end{split}
\end{equation}
Measuring deviations in the SGWB anisotropies at low $\ell$ from the angular power-spectrum for $f_{\rm dec}(\eta_i)=0$ would be a proof that at early epochs there were decoupled relativistic  particle species contributing to the total energy density by an amount $f_{\rm dec}(\eta_i)$. Notice that $\eta_i$ would correspond to temperatures so high that any Standard Model particle would be coupled at that epoch, thus any species responsible for this effect should arise in theories beyond the Standard Model. We can therefore conclude that this damping effect would provide precious information about new physics.
The most promising candidate which can give important contributions to $f_{\rm dec}(\eta_i)$ is extra-radiation (ER), parametrized as $\Delta N_{\rm eff}$, the excess from the standard value of 3.046 for the effective neutrino number $N_{\rm eff}$~\cite{Calabrese:2011hg}. These new species are relativistic at the present epoch, so they were relativistic at the end of inflation too. They cannot be Standard Model particles, therefore it is reasonable to suppose that they decoupled at temperatures higher than the energies reached in modern accelerators, $T_{dec}^{ER}\gtrsim 10^6  {\rm GeV}$. This is consistent with the hypothesis that at the end of inflation they were decoupled too, i.e. $\eta_{T_{dec}^{ER}}\lesssim \eta_i$. On the other hand, if we fix a specific particle physics model, we are able to describe the evolution of the decoupled relativistic  degrees of freedom, or, in other words, we know $f_{\rm dec}(\eta)$. 
Under such a hypothesis, a measurement of the SGWB angular power-spectrum would allow to determine a range for $\eta_i$, on the basis of the evolution of $f_{\rm dec}$.

\begin{figure}[t!]
\centering 
\includegraphics[width=0.5\textwidth]{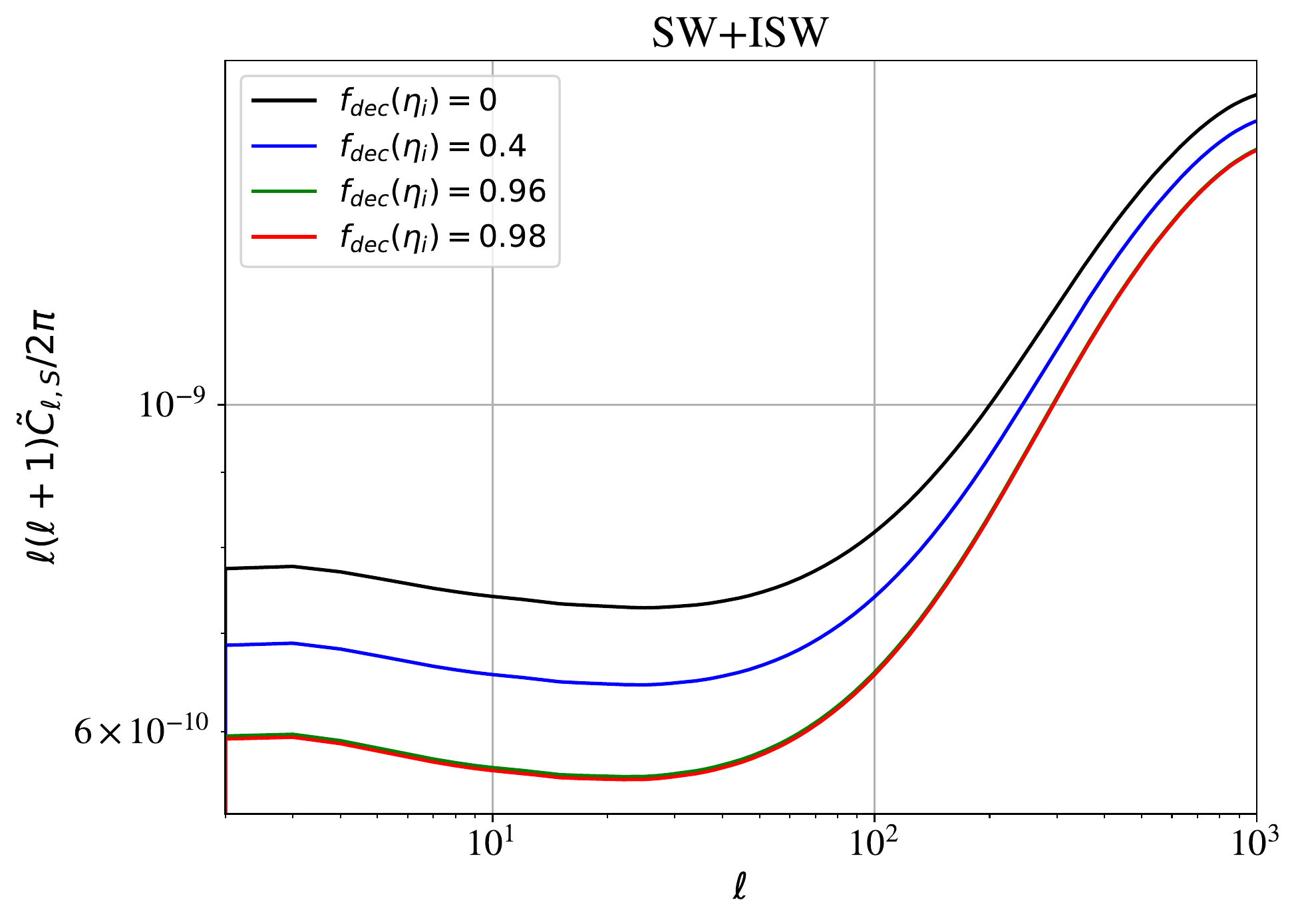}
\vskip -0.4cm
\caption{\it Total scalar contribution to the SGWB angular power-spectrum, sum of the SW and the ISW terms.}
\label{fig:swisw}
\end{figure}

\paragraph{Integrated Sachs-Wolfe effect and relativistic particle species. }
As anticipated, the ISW effect is roughly proportional to the total variation of the potentials $\Delta \Phi+\Delta\Psi$, so, when we consider the total variation for the SGWB, we end up with larger variations with respect to the CMB, because in the CMB case we take the difference between the initial value (at recombination) and the present epoch, but at recombination the potentials were already damped, especially at small scales, therefore they would have a smaller impact on the ISW. 
\begin{figure}[t!]
\centering 
\includegraphics[width=0.48\textwidth]{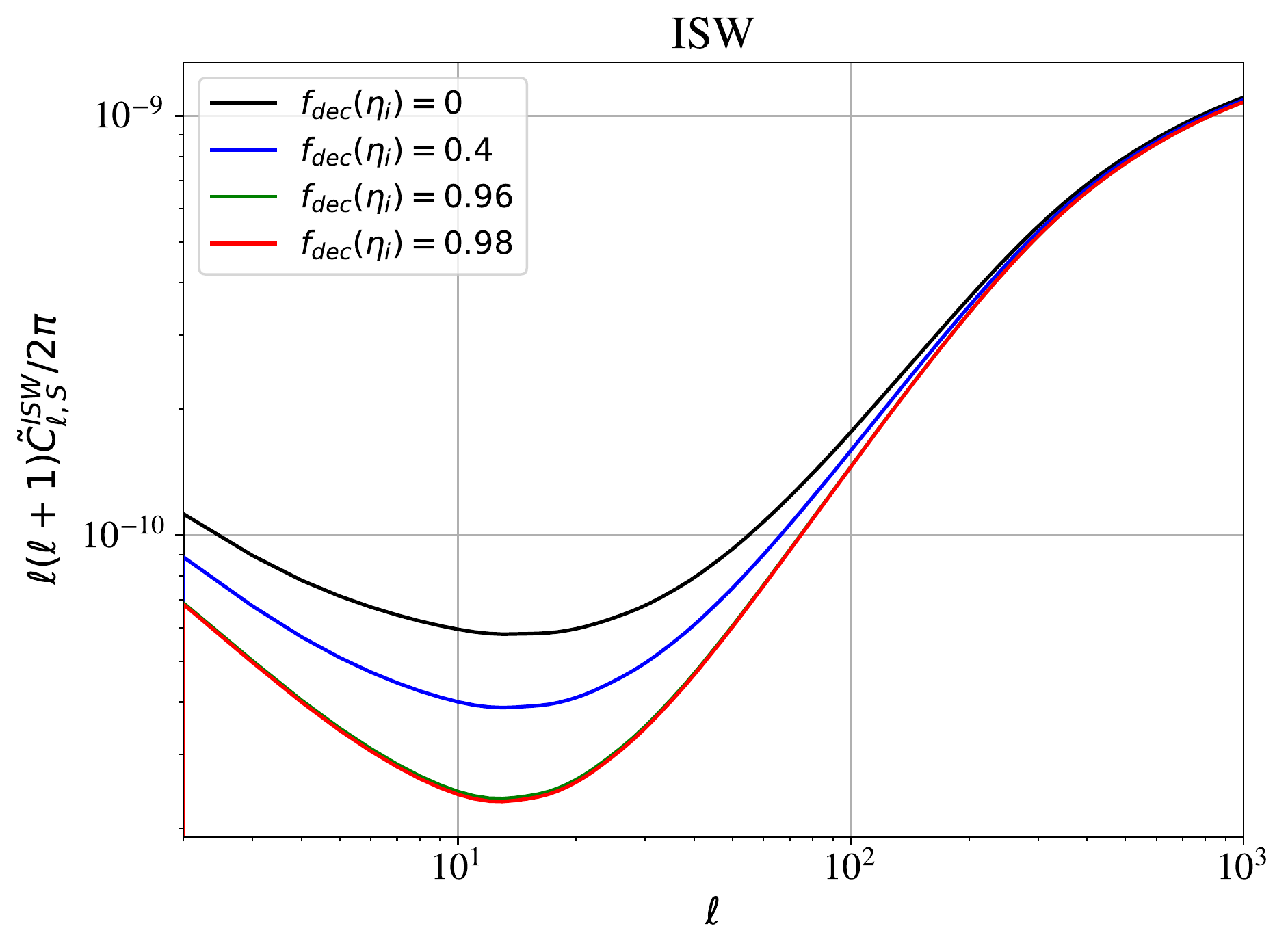}
\vskip -0.4cm
\caption{\it ISW contribution to $\Tilde{C}_{\ell,S}$. We observe a bump at large $\ell$ due to the fact that the potentials at large $\ell$ have the maximum variation.}
\label{sgwb_isw_only}
\end{figure}
The ISW contribution depends upon the variation of the potentials between  $\eta_i$ and  $\eta_0$; so it is sensitive the evolution of $f_{\rm dec}(\eta_i)$ up to  low energies scales $(T\lesssim0.1$ MeV). Thus measurements of the anisotropies of the SGWB anisotropies can constrain extra particles species both at high and low energy scales.
The effect of the change of number of relativistic degrees of freedom on the ISW contribution to the angular power-spectrum is represented in Fig.~\ref{sgwb_isw_only}. As anticipated, a higher number of relativistic species suppresses the ISW contribution at the largest angular scales through its effect on the early ISW contribution.\\
We can sum up the two ``scalar'' contributions to show the main effect on large angular scales that, in the future, can be probed by GW interferometers. The result is given in Fig.~\ref{fig:swisw}, 
where  the impact of a varying number of decoupled relativistic species is evident.
We did not consider the contribution coming from the tensor background perturbations since we checked that they do not alter the 
spectrum at scales that can be probed in the future by GW direct detection experiments. 
As well known~\cite{Weinberg:2003ur} decoupled relativistic particles, and in particular neutrinos, create a damping on the amplitude of the tensor modes in the CMB. In a similar way relativistic particles have an impact also on the monopole amplitude of the GW energy density~\cite{Watanabe:2006qe, Mangilli:2008bw} and so on the amplitude of the angular power-spectrum.

\paragraph{Conclusions.} 

\noindent In this Letter we have shown that the future detection of the SGWB of cosmological origin has profound implications on our understanding of the physics of the early Universe and on high energy physics aspects not accessible by present-day particle accelerators. We have shown that the anisotropies of the SGWB inherited by the GW generated during their propagation in the Universe, from the time of their decoupling at the end of inflation until today, feel the effect of relativistic particle species that are decoupled from the thermal bath. Having in mind the poor angular resolution of future GW detectors, we have focused on the effects most relevant at very large scales. As for the CMB, also for the SGWB, such scales are affected by the SW effect and by the ISW effect. We have therefore quantified the effect of different particle species on both the SW and ISW, and we have computed the SGWB angular power-spectrum.
The cumulative effect of a larger number of decoupled relativistic  particle species on the angular power-spectrum of the SGWB is a suppression at large scales. 
This will clearly becomes a potential observable effect as soon as such anisotropies will be detected.

\vskip .2cm


\vskip .1cm

\paragraph{Acknowledgments.}
\noindent
N.B., D.B. and S.M. acknowledge partial financial support by ASI Grant No. 2016-24-H.0.

\end{document}